\documentclass[a4paper]{spie}  
\usepackage[]{graphicx}

\usepackage[left=1.925cm, right=1.925cm, top=2.54cm, bottom=4.94cm]{geometry}

\title{Prospects for near-infrared characterisation of hot Jupiters with the VLTI Spectro-Imager (VSI)} 

\author{S. Renard\supit{a}, O. Absil\supit{a}, J.-P. Berger\supit{a}, X. Bonfils\supit{b}, T. Forveille\supit{a}, F. Malbet\supit{a}
\skiplinehalf
\supit{a}Laboratoire d'Astrophysique de l'Observatoire de Grenoble (LAOG), France; \\
\supit{b}Observatorio Astronomico de Lisboa, Lisboa, Portugal
}

\authorinfo{Further author information: (Send correspondence to
  S.R.) \\ S.R.: E-mail: Stephanie.Renard@obs.ujf-grenoble.fr,}

\DeclareTextSymbol{\degre}{OT1}{23}

\begin{document} 
\maketitle 

\begin{abstract}
In this paper, we study the feasibility of obtaining near-infrared spectra 
of bright extrasolar planets with the 2nd generation VLTI 
Spectro-Imager instrument (VSI), which has the required angular
resolution to resolve nearby hot Extrasolar Giant Planets (EGPs) from their host stars. Taking
into account fundamental noises, we simulate closure phase measurements
of several extrasolar systems using four 8-m telescopes at the VLT and
a low spectral resolution ($R = 100$). Synthetic planetary spectra from
T. Barman are used as an input. Standard $\chi^2$-fitting methods are
then used to reconstruct planetary spectra from the simulated data.
These simulations show that low-resolution spectra in the $H$ and $K$ bands
can be retrieved with a good fidelity for half a dozen targets in a
reasonable observing time (about 10 hours, spread over a few nights).
Such observations would strongly constrain the planetary temperature
and albedo, the energy redistribution mechanisms, as well as the
chemical composition of their atmospheres. Systematic errors, not
included in our simulations, could be a serious limitation to these
performance estimations. The use of integrated optics is however
expected to provide the required instrumental stability (around $10^{-4}$ on the closure phase)
to enable the first thorough characterisation of extrasolar planetary emission
spectra in the near-infrared.
\end{abstract}



\section{INTRODUCTION}
\label{sec:intro}  

Since the discovery by Mayor \& Queloz\cite{mayor95} of the first exoplanet around 51 Pegasi, the study of planetary systems has received an increasing attention, with the continuous development of new techniques. Among the direct detection techniques, interferometry is one of the most promising for the near future. It already provides the required angular resolution, but the dynamic range needs to be improved. The detection and characterization of extrasolar planets is one of the main science cases of the 2nd generation VLTI Spectro-Imager instrument (VSI)\cite{VSIMalbet}.

\begin{figure}[tp]
  \begin{center}
    \begin{tabular}{cc}
      \includegraphics[width=0.45\textwidth]{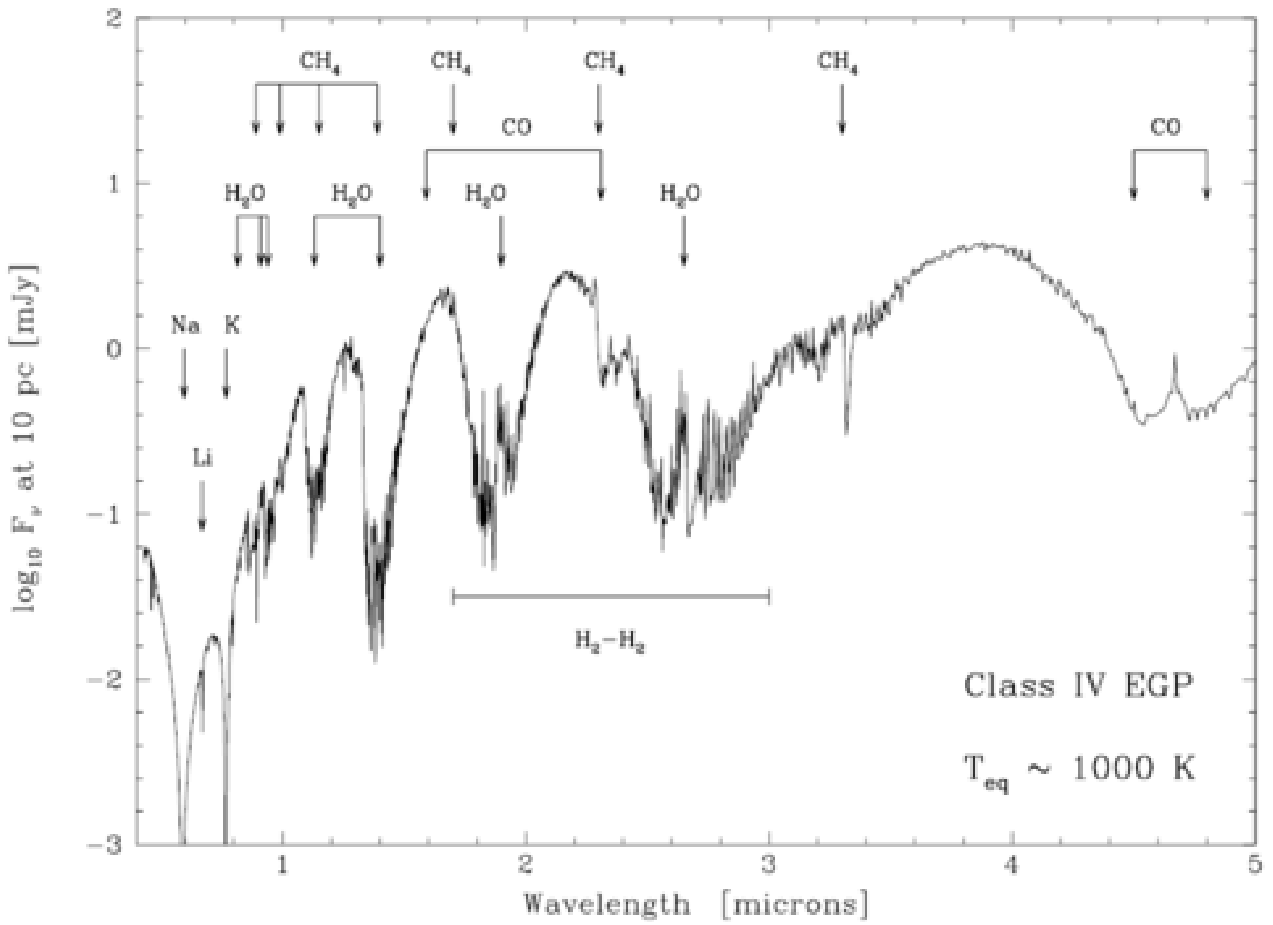}&
      \includegraphics[width=0.45\textwidth]{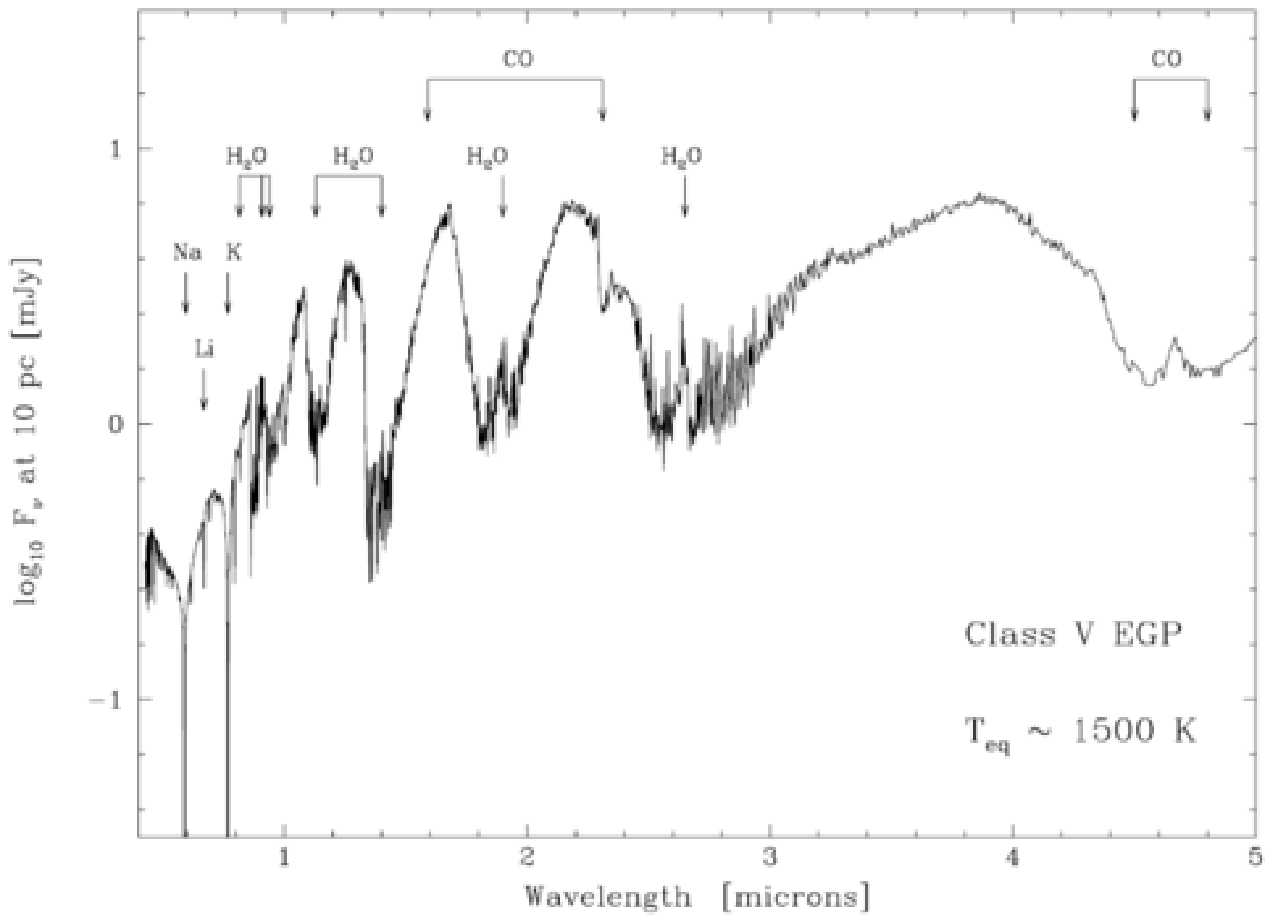}
    \end{tabular}
  \end{center}
  \caption[step] 
          { \label{fig:sudarsky} 
            Model fo spectra for hot EGPs, with $T_{eq} = 1000$K and 1500K, orbiting a Sun-like star. Important changes in abundances and band strength are seen between both cases\cite{sudarsky03}
            }
\end{figure}

At a distance of $a \approx 0.05$ AU, most of the detected exoplanets, called hot Extrasolar Giant Planets (EGPs), receive from their parent star about $10^4$ times the amount of radiation intercepted by Jupiter from our Sun. So close to the star, they are also believed to be tidally locked such that half of the planet faces permanently the star while the other half stays in the dark. The fraction of incident light absorbed by the atmosphere -- parametrized by its albedo $A$ -- heats the planets. This heat is redistributed to the night side by strong wind and reradiated by the planet.
Besides providing their heat source, the strong radiation illuminating hot EGPs also structures their atmosphere. It suppresses the convection to depths well below the photosphere, leading to a fully radiative photosphere across most of the day side\cite{guillot96}. Whether silicate clouds can persist in the photosphere then results from the competition between the sedimentation and advective timescales. The sedimentation timescale of radiative photosphere is short and the winds are believed not to be strong enough to prevent dust from settling. Hence, the atmospheres of hot EGPs are pictured as being free of clouds. 
Finally to reproduce hot-Jupiter spectra, their chemical composition should be taken into account. As the opacity of each species regulates the emergent flux as a function of wavelength, their emergent spectra strongly depart from an ideal black body assumption. They display large molecular bands and spectral features (see Fig.~\ref{fig:sudarsky}).

The transit technique revolutionised the field of exoplanetology. It was a breakthrough to peer into hot EGPs' structure and to already give a glimpse about their composition. However, the error bars remain large to discriminate between different models and most of the planetary spectrum remains unknown. Particulary, the $1 - 2.4$ $\mu$m spectrum is very rich in spectral information and may provide unprecedented constraints on our understanding of planetary atmosphere. VSI will have the ability to observe hot EGPs in $J$, $H$ and $K$ bands, with a low ($ R = 100$) or medium ($R = 1000$) resolution. Model fitting of low-resolution spectra gives a measurement of their albedo and test the cloud-free assumption. The phase dependance of the measured signal shall constrain the heat redistribution (through the temperature across the surface) and the weather conditions. At medium resolution, it will be possible to measure the abundance of CO and test the presence of CH$_4$. Contrarily to current characterisation techniques, VSI will not be restricted to transiting planets. The sample of favourable targets counts already 7 planets today and may count at least twice this number at the time of VSI will be in operation. Therefore, VSI will not only enhance our knowledge on few transiting planets, it will also allow statistical studies to be carried out and literally enable comparative exoplanetology.

The goal of this work is to study the feasibility of obtaining near-infrared emission spectra of bright extrasolar giant planets (EGPs) with VSI. In Sect.~\ref{sec:meth}, we explain the method used for the simulation, from the choice of the targets to the simulations. Sect.~\ref{sec:res} contains the results of the simulation, and Sect.~\ref{sec:conclu} the discussion and the conclusions of the study.

\section{METHOD}
\label{sec:meth}

\begin{figure}[tp]
  \begin{center}
    \begin{tabular}{c}
      \includegraphics[width=0.5\textwidth]{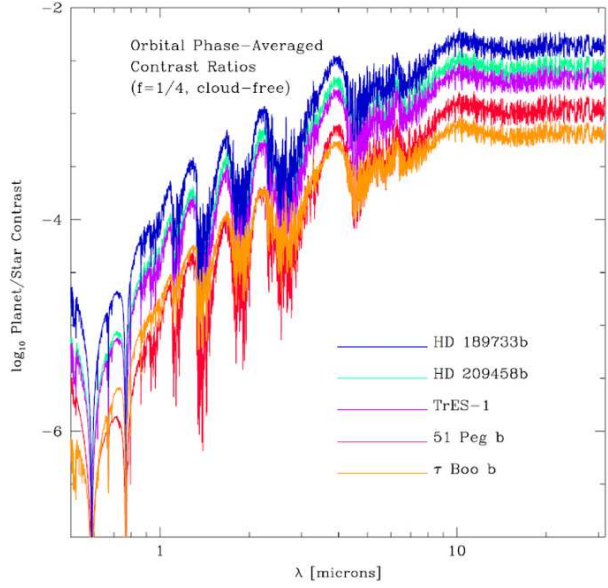}
    \end{tabular}
  \end{center}
  \caption[step] 
          { \label{fig:spectre} 
            Planet-to-star contrast ratio for few favourable exoplanet\cite{2005ApJ...625L.135B}
          }
\end{figure}

In the following simulations,  we use the synthetic spectra developped by Burrows\cite{2001ApJ...556..885B} in order to assess the feasability of planetary spectra characterisation with VSI. The latest models for the brightest and closest hot EGPs have been kindly provided by T. Barman and are illustrated in Fig.~\ref{fig:spectre}.

\subsection{Choice of targets}

To determine the feasibility of EGP spectroscopy with VSI, we started to simulate interferometric observations of several EGPs that have been discovered by radial velocity surveys. The suitable targets for an interferometric study are the hot extrasolar giant planets that orbit close to their parent stars, for which the star/planet contrast does not exceed a few $10^4$ in the near-infrared. Another criterion is that the target must be close enough so that VSI, with its angular resolution of a few milli-arcsec, can resolve the star-planet system. Typically, hot Jupiter systems further than 50 pc fall outside the resolving power of the VLTI. Finally, we have restricted the list target to declinations ranging between -84\degre and +36\degre, in order to be observable from Cerro Paranal.

A list of six targets, based on these criteria, has been compiled in Table~\ref{tab:caract1}. The properties of the planetary companions are listed in Table~\ref{tab:caract2}. These targets were expected to give better results in terms of detectability. In order to carry out performance simulations we have used the synthetic spectra computed by Barman\cite{2001ApJ...556..885B}.

\begin{table}[tp]
\caption{
  Parameters of the host stars for the selected extrasolar planetary
  systems\cite{2006ApJ...646..505B,Schneider07}. The planetary radii
  are derived from transit measurements when available, while the
  values followed by an asterik are estimates using a mean planet
  density\cite{Bouchy05} of 0.7 g cm$^{-3}$ and a
  upper limit of 1.5R$_J$. The semi-major axis is given in AU and in
  mas, assuming the planet to be at maximum elongation. The estimate
  temperature and flux of the hot EGPs are computed using a grey
  body assumption with a Bond albedo of 0.1. The stellar and planetary
  fluxes, as well as the planet/star contrast, are given in the centre
  of the $K$ band.}
\label{tab:caract1}
\scriptsize
  \begin{center}
    \begin{tabular}{ccccccccccc}
      \hline
      \hline
      Star & RA & Dec & m$_V$ & m$_J$ & m$_H$ & m$_K$ & Distance & Spectral & Star &
      Apparent \\
      name & (J2000) & (J2000) & & & & & (pc) & type & radius (R$_{\odot}$) & diameter (mas) \\
      \hline
      $\tau$ Boo & 13h47m17s & +17\degre27'22'' & 4.5 & 3.6 & 3.5 & 3.5 &
      16 & F7V & 1.43 & 0.43  \\
      HD 179949 & 19h15m33s & -24\degre10'45'' & 6.3 & 5.3 & 5.1 & 4.9 &
      27 & F8V & 1.23 & 0.21 \\
      HD189733 & 20h00m44s & +22\degre42'39'' & 7.7 & 6.1 & 5.6 & 5.5 & 19
      & K1-2V & 0.777 & 0.19 \\
      HD 73256 & 08h36m23s & -30\degre02'15'' & 8.1 & 6.7 & 6.4 & 6.3 & 37 &
      G8/K0 & 0.99 & 0.13 \\
      51 Peg & 22h57m27s & +20\degre46'07'' & 5.5 & 4.7 & 4.2 & 3.9 &
      15 & G2IV & 1.15 & 0.36 \\
      HD 209458 & 22h03m10s & +18\degre53'04'' & 7.7 & 6.6 & 6.4 & 6.3
      & 47 & G0V & 1.15 & 0.11 \\
      \hline
    \end{tabular}
  \end{center}
\end{table}

\begin{table}[tp]
\caption{
  Parameters of the selected extrasolar
  planets\cite{2006ApJ...646..505B,Schneider07}} 
\label{tab:caract2}
  \begin{center}
    \begin{tabular}{cccccccccc}
      \hline
      \hline
      Planet & M $\sin i$ & Radius & Temp & axis & axis & period &
      F$_*$ & F$_p$ & Contrast \\
      name & (M$_J$) & (R$_J$) & (K)  & (AU)  & (mas)  & (days)  &
      (Jy) & (Jy) & ($K$ band) \\
      \hline
      $\tau$ Boo b & 3.9 & 1.5* & 1607 & 0.046 & 2.95 & 3.31 & 27.3 &
      8.1e-3 &  3.0e-4  \\
      HD 179949 b & 0.92 & 1.2* & 1533 & 0.045 & 1.67 & 3.09 & 6.61 &
      2.2e-3 & 3.4e-4 \\
      HD189733 b & 1.15 & 1.16 & 1180 & 0.031 & 1.61 & 2.22 & 3.60 &
      9.4e-3 & 2.6e-4 \\
      HD 73256 b & 1.87 & 1.5* & 1296 & 0.037 & 1.01 & 2.55 & 1.82 &
      4.3e-3 & 2.3e-4 \\
      51 Peg b & 0.47 & 0.9* & 1265 & 0.052 & 3.54 & 4.23 & 17.3 &
      2.3e-3 & 1.3e-4 \\
      HD 209458 b & 1.32 & 1.32 & 1392 & 0.045 & 0.96 & 3.52 & 1.81 &
      4.8e-3 & 2.6e-4 \\
      \hline
    \end{tabular}
  \end{center}
\end{table}

\subsection{The method : differential closure phases}

Because hot EGPs are very close to their parent star, they are the planets with both the largest source of heat and the largest reflected starlight. In brief, they are the most luminous amoung known exoplanets. Nevertheless, they remain very difficult targets for direct observations. The typical contrast between hot EGPs and their parent star ranges from $10^{-6}$ in the visible to $10^{-3}$ beyond $10$ $\mu$m while typical angular separations are of the order of 1 mas. In the wavelength domain of VSI, common hot EGPs have a typical contrast of $10^{-4}$ and the best targets reach a contrast of $10^{-3}$ (see Fig.~\ref{fig:spectre}).

Observations at such contrast are challenging and different strategies are currently investigated. The resolving power of interferometry provides the means to achieve observations of hot EGPs and two approaches have been investigated: differential phase and differential closure phase. The first approach measures the photo-center of the planet-star system (as a function of wavelength) whereas the second measures the fraction of light  that is not point-symmetric (as a function of wavelength too). In both cases, it provides the differential planet to star contrast ratio as a function of wavelength.

\begin{figure}[tp]
  \begin{center}
    \begin{tabular}{c}
      \includegraphics[width=0.6\textwidth]{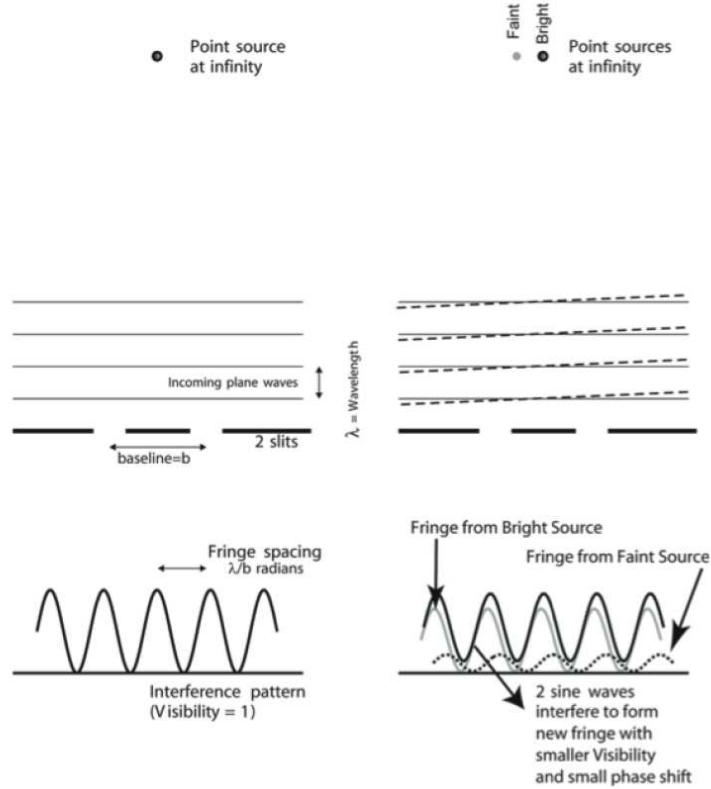}
    \end{tabular}
  \end{center}
  \caption[step] 
          { \label{fig:beuzit} 
             The figure illustrates how the presence of a planet modifies the phase of interferometric fringes\cite{beuzit07}            
          }
\end{figure}

\begin{figure}[tp]
  \begin{center}
    \begin{tabular}{c}
      \includegraphics[width=0.6\textwidth]{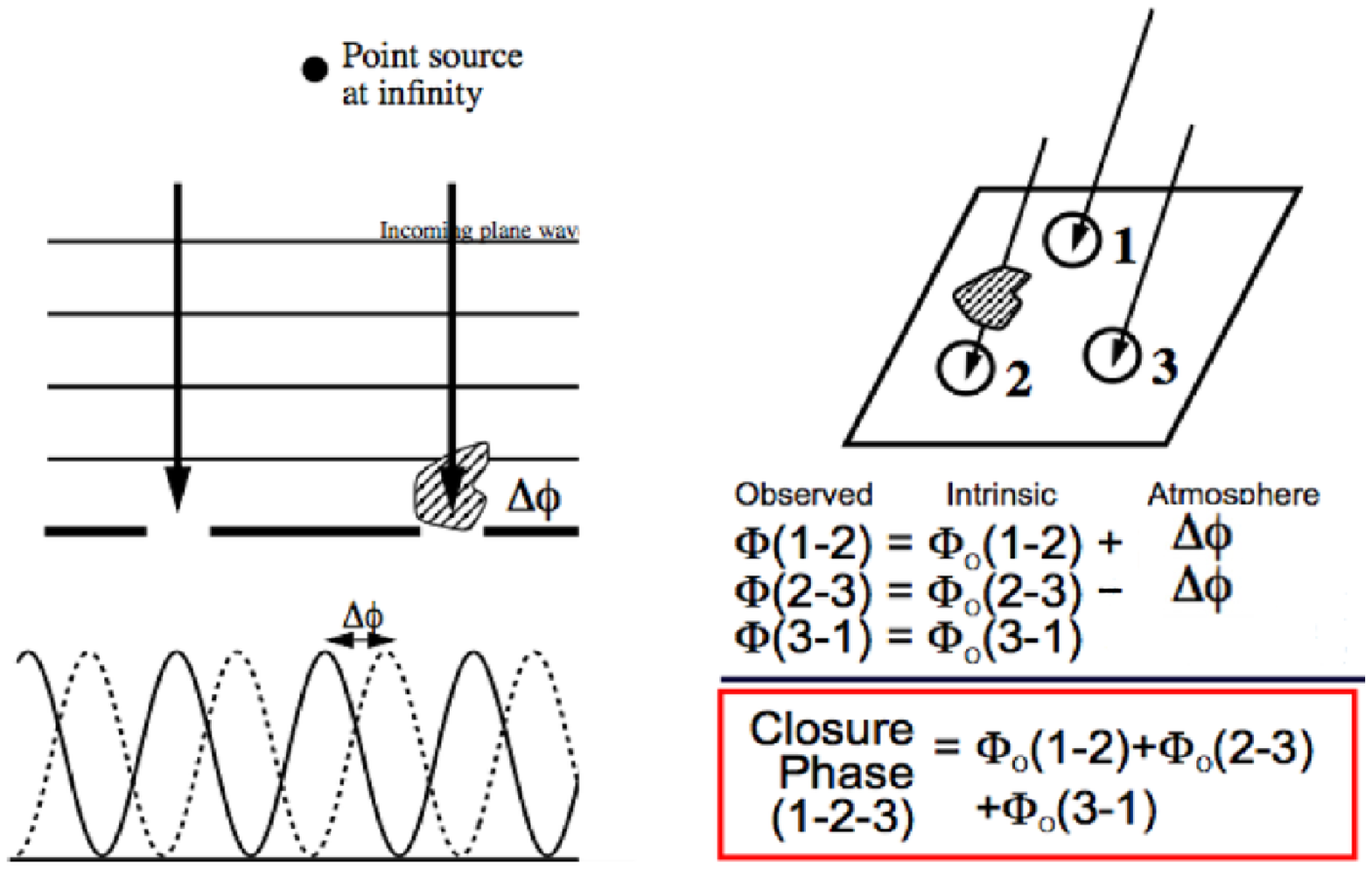}
    \end{tabular}
  \end{center}
  \caption[step] 
          { \label{fig:CPExpl} 
            The figure illustrates how a phase delay introduced above
            a telescope can be canceled in the closure phase
            quantity. The closure phase equals the sum of phases
            measured along baselines formed by at least three telescopes\cite{Lawson00}.
            }
\end{figure}

When a star comes with a faint companion and when both fall in the field of view of an interferometer, their fringe patterns sum together incoherently. The presence of a planet decreases the fringe contrast and changes the phase by tiny amounts (see Fig.~\ref{fig:beuzit}).
From the ground, the main problem is the Earth atmosphere, where transmission varies chromatically and on time scales shorter than the time required to perform the observations. Such wavelength-dependent phase shifts essentially prevent the use of the ``differential phase'' technique for high contrast observations because this technique requires therefore an extremely good control and calibration of the atmospheric and instrumental stability\cite{2006MNRAS.367..825V,2003EAS.....8..297P,2000SPIE.4006..407L}. However, with three or more telescopes, one can build an interferometric observable that is robust to phase shifts: the closure phase, which presents the nice property of cancelling atmospheric systematics. As shown in Fig.~\ref{fig:CPExpl}, a differential optical path above one telescope (here the $2^{nd}$) introduces phase shifts on the fringes measured on two baselines (here baselines $b_{1-2}$ and $b_{2-3}$). Because, the phase shifts have opposite signs, they cancel out when summed together. Since the reasoning holds for a phase delay introduced above any of the telescope, the sum of phases measured on baselines  $b_{1-2}$,  $b_{2-3}$ and  $b_{1-3}$ cancels all telescope-based phase systematics. Actually, the closure phase technique does not only cancel the phase shifts introduced by the atmosphere but also the ones introduced by the instrumental optics, up to the recombination. 

Therefore, differential closure phase appears to be one of the best methods to obtain medium-resolution near-infrared spectra of known hot EGPs, because it is very sensitive to faint companions while not corrupted by random atmospheric phase fluctuations.

\subsection{Steps of the simulation}

\begin{figure}[tp]
  \begin{center}
    \begin{tabular}{cc}
      \includegraphics[width=0.4\textwidth]{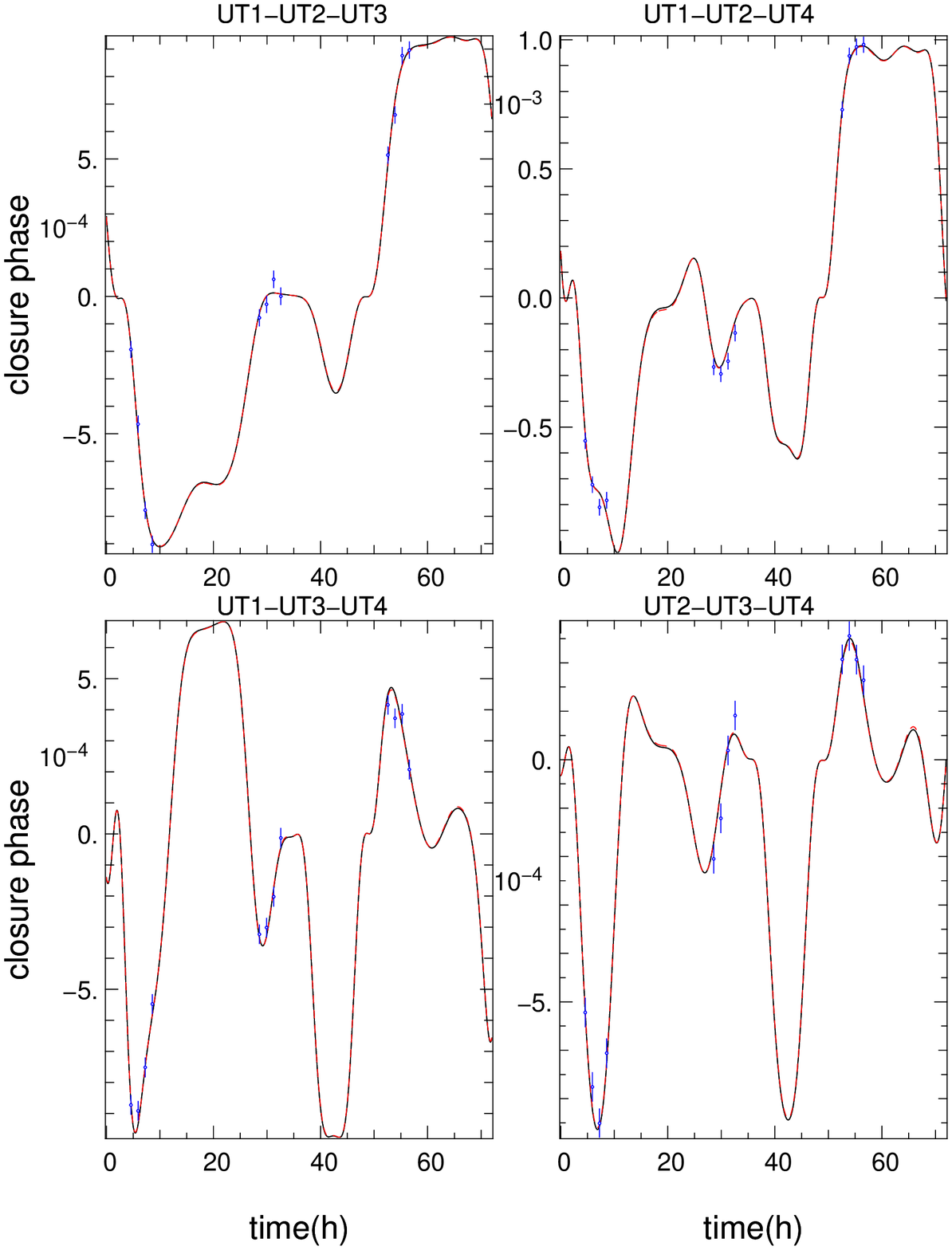}&
      \includegraphics[width=0.4\textwidth]{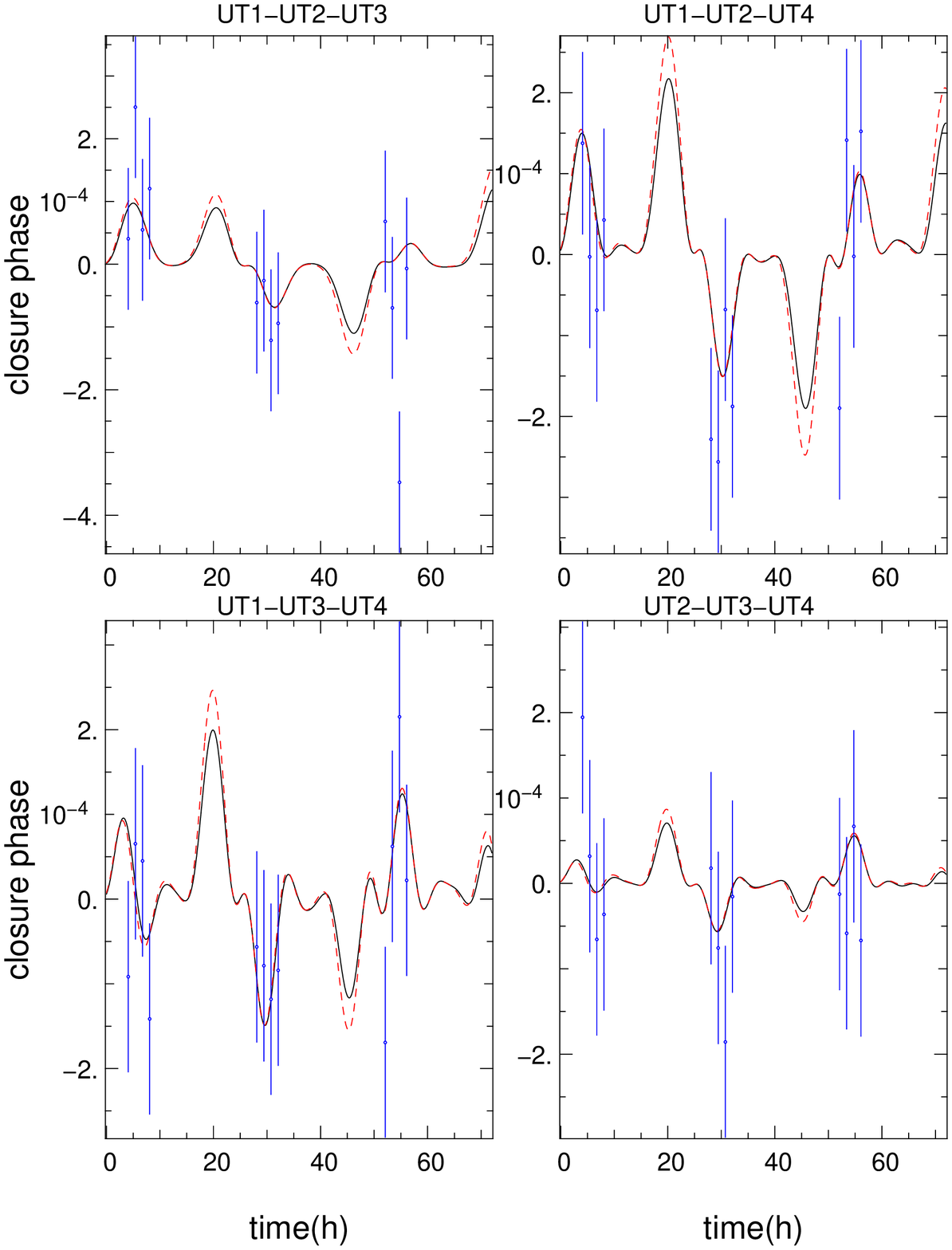}
    \end{tabular}
  \end{center}
  \caption[step] 
          { \label{fig:step_simu} 
            Simulated observations of two extrasolar planetary systems
            (left : $\tau$ Boo, right : HD 73256) using VSI with 4
            UTs, hence 4 triangles. The data points with their associated bars (in blue)
            are shown separately for the four triplets of
            baselines. The underlying black curve represents the
            noiseless closure phase signal of the extrasolar planet
            for each triplet, while the red curve represents the best
            fit to the simulated data, as described in
            Sect.~\ref{subsect:fit_param_orbit}. The simulations are
            performed in the band K.
            }
\end{figure}

To perform the detection of exoplanets from closure phase data, we rely on model fitting. Image reconstruction is not needed in our case, as the system of a hot Jupiter plus a star is just the case of a binary, though a highly contrasted one. The performance simulations will therefore consist in the following steps:
\begin{enumerate}
\item To simulate the observation of a star-planet system under typical
  conditions : we assume that the system is observed on three
  consecutive nights with the four UTs and that four data points are
  acquired each night at a rate of one data point per hour. Each data
  point consist in a 10-min on-source integration, and is followed by
  calibration measurements (not simulated here). 
\item Estimate the fundamental noises (shot noise and detector noise)
  for each individual measurement, using the lowest spectral resolution of VSI
  ($R=100$). The error bars on individual data points typically range
  between  $3 \times 10^{-5}$ and $10^{-4}$ radians for stellar magnitude
    between 3.5 and 6.5 in the $K$ band. Using these estimated
    error bars, we draw random data points using a gaussian distribution
    centered around the noiseless closure phase and with a standard
    deviation equal to the error bar. This results in a collection of data
    points with associated error bars (see
    Fig.~\ref{fig:step_simu}), which are used as inputs for the
    fitting procedure.
\item Fit the simulated observations and their associated error bars
    with a binary model for the closure phase of the planetary system as a
    function of time. The time evolution of the closure phase
    simultaneously captures the motion of the star-planet system on the
    night sky and the orbit of the planet around its host star
    (typical period of 3 days for the systems considered here).
\end{enumerate}

\section{Results}
\label{sec:res}

\subsection{Determination of the orbital parameters}
\label{subsect:fit_param_orbit}

Various free parameters can be used to perform the fit of the simulated closure phase observations. Here, we select the three most important parameters that are not known from radial velocity measurements : the planet/star contrast, the orbital inclination and the position angle of the orbit on the plane of the sky\footnote{The latter is counted East of North and is actually equivalent to the longitude
  of the ascending node of the orbit with respect to the plane of the
  sky.}. In a first stage, we fit the three parameters globally on the
whole spectral domain. Because the contrast significantly changes
between individual spectral channels, we replace the first fitting
parameter (the contrast) by the planetary radius. Fitting the
planet/star contrast is indeed equivalent to fitting a planetary
radius in each channel if one makes the following assumptions : 
\begin{itemize}
\item The thermal emission follows a grey body emission law.
\item The albedo is constant and fixed to given value (0.1 in our
  case).
\item The temperature of the planet is computed from radiative equilibrum.
\end{itemize}

It must be noted that the choice to compute the fit on the planet radius rather than on the albedo is due to the fact that, when fitting the data, the albedo changes rapidly for small variations of the radius and can quickly reach non-physical values if a poor estimation of the radius is chosen. This come from $F_{thermal} \propto (1-A_b)^{1/4}R^2_p$, so that a variation of $R_p$ has a bigger influence on $F_{thermal}$ than a variation of $A_b$.

\begin{figure}[tp]
  \begin{center}
    \begin{tabular}{cc}
      \includegraphics[width=0.4\textwidth]{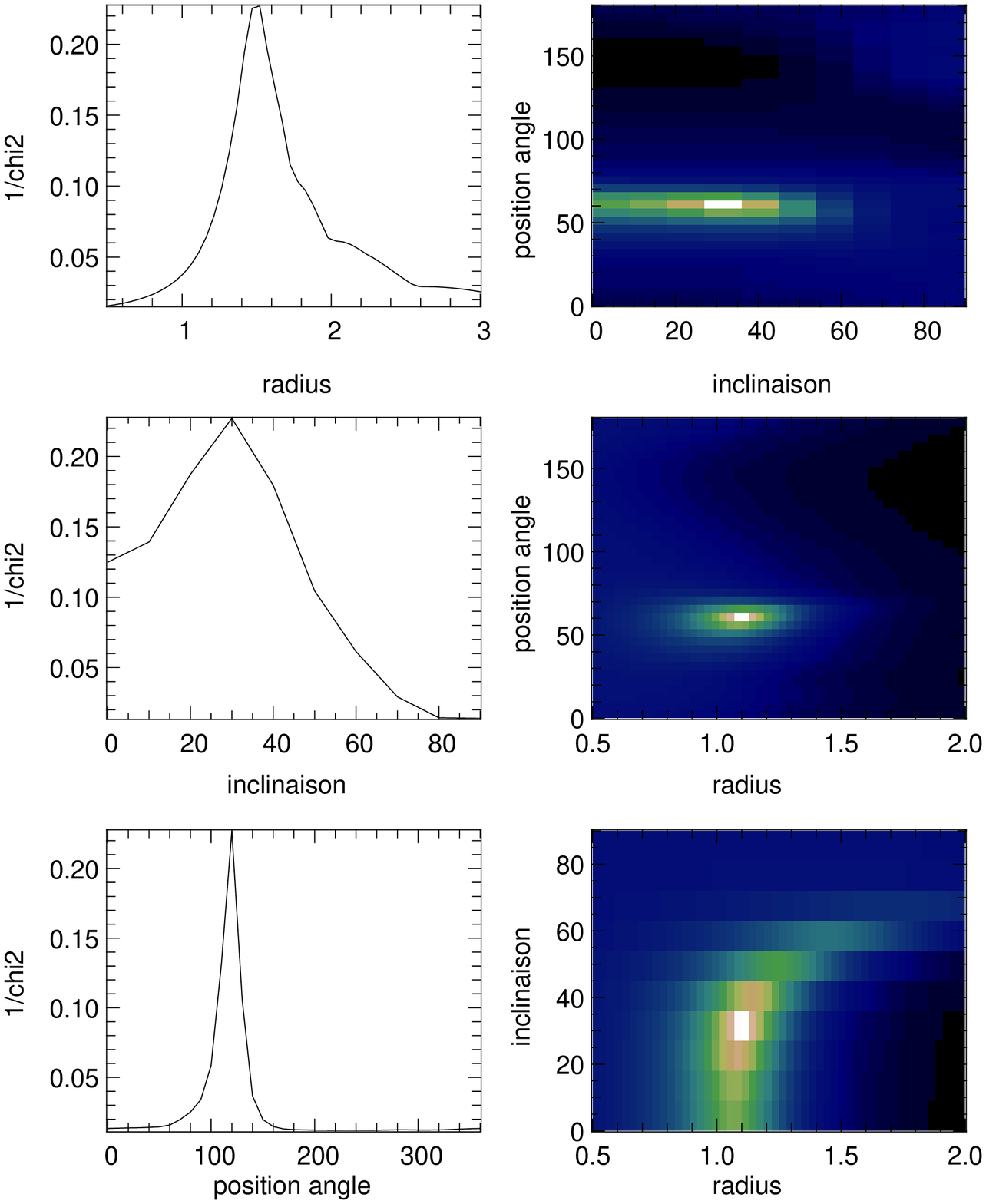}&
      \includegraphics[width=0.4\textwidth]{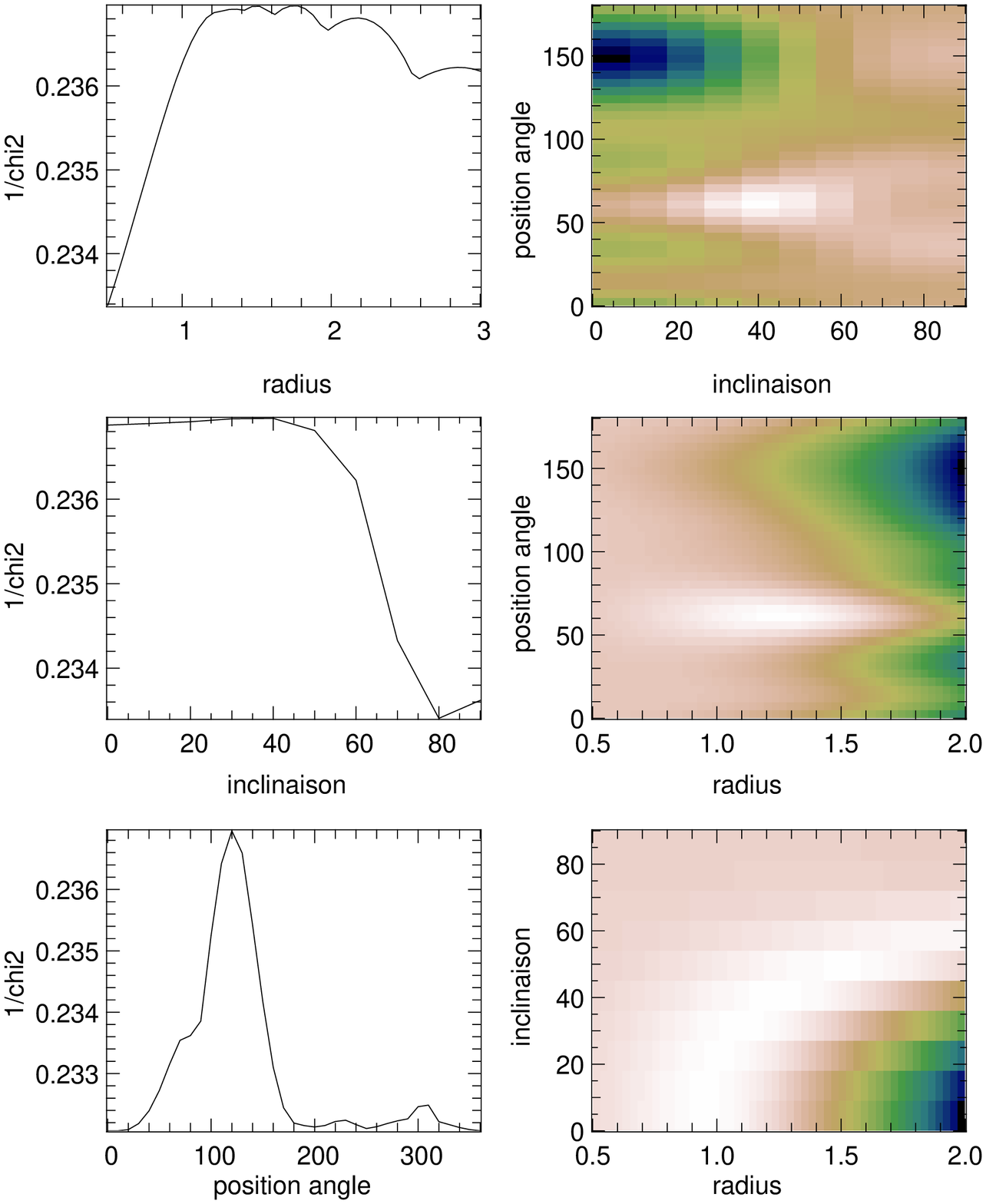}
    \end{tabular}
  \end{center}
  \caption[step] 
          { \label{fig:fit_chi2} 
            Graphical representations of the $\chi^2$ cube between the simulated observations and the model for a whole range of values for the three free parameters : the radius, the inclination and the position angle
            (left : $\tau$ Boo, right : HD 73256).
            }
\end{figure}

The fit is performed in two successive steps. In a first step, we
investigate the whole parameter space and compute the $\chi^2$ between
the observations and the model for a whole range of values for the
radius, inclination and position angle. From the resulting $\chi^2$
hypersurface, we determine an approximate position for the global
minimum, which we will use as an initial guess in the second step of
the fit. The graphical representations of the $\chi^2$ cube (see Fig.~\ref{fig:fit_chi2}) allow us to evaluate the sharpness of the $1/\chi^2$ peak and the possible occurence of local minima. The second step consists in a classical Levenberg-Marquardt
$\chi^2$ minimisation with three free parameters. This step
usually converges quickly towards the best-fit solution, as the
initial guess is generally robustly determined during the first step
of the fitting procedure.

While the output best-fit radius does not have a real physical meaning
under the present asumptions (grey body with fixed albedo and
temperature), the inclination and position angle of the orbit are
generally well reproduced by the fitting procedure, both in the $H$ and
in the $K$ band (see Tables~\ref{tab:Results_fit_KBand} \&
\ref{tab:Results_fit_HBand}). These two
parameters are generally unknown for the simulated planets, so that
arbitrary values have been used in this study.

\begin{table}[tp]
\caption{Results of the orbital fit for the 6 selected exoplanets
      in $K$ band}
   \label{tab:Results_fit_KBand}
    \begin{center}
       \begin{tabular}{ccccccc}
        \hline
        \hline
            & $\tau$ Boo b & HD 179949 b & HD 189733 b & HD 73256 b &
        51 Peg b & HD 209458 b \\
        \hline
          inclination &  30 & 45 & 85.76 & 30 & 75 & 86.929 \\
          best fit   & 28.74 & 50.02 & 70.13 & -4.83 & 75.28 & 87.86  \\
          error bar & 0.63 & 5.67 & 11.08 & 326.29 & 2.29 & 6.91 \\
        \hline
          position angle & 120 & 60 & 90 & 120 & 150 & 0\\
          best fit & 119.87 & 59.91 & 96.78 & 110 & 150.13 & 336.09 \\
          error bar & 0.36 & 2.19 & 6.53 & 10.37 & 2.10 & 7.73 \\
        \hline
      \end{tabular}
   \end{center}
\end{table}

\begin{table}[tp]
    \caption{Results of the orbital fit for the 6 selected exoplanets
      in $H$ band}
    \label{tab:Results_fit_HBand}
\begin{center}
       \begin{tabular}{ccccccc}
        \hline
        \hline
            & $\tau$ Boo b & HD 179949 b & HD 189733 b & HD 73256 b &
        51 Peg b & HD 209458 b \\
        \hline
          inclination &  $30$ & $45$ & $85.76$ & $30$ & $75$ & $86.929$ \\
          best fit & 30.47 & 28.60 & 98.67 & 43.22 & 45.93 & 80.05  \\
          error bar & 0.56 & 11.31 & 25.61 & 21.61 & 8.72 & 14.17 \\
        \hline
          position angle & 120 & 60 & 90 & 120 & 150 & 0\\
          best fit & 119.62 & 59.33 & 87.21 & 122 & 156.27 & 331.07 \\
          error bar & 0.43 & 2.15 & 11.28 & 7.38 & 6.50 & 9.96 \\
        \hline
      \end{tabular}
\end{center}
\end{table}

\subsection{Determination of the planetary spectra}

\begin{figure}[tp]
  \centering
  \begin{tabular}{ccc}
    \includegraphics[width=0.3\textwidth]{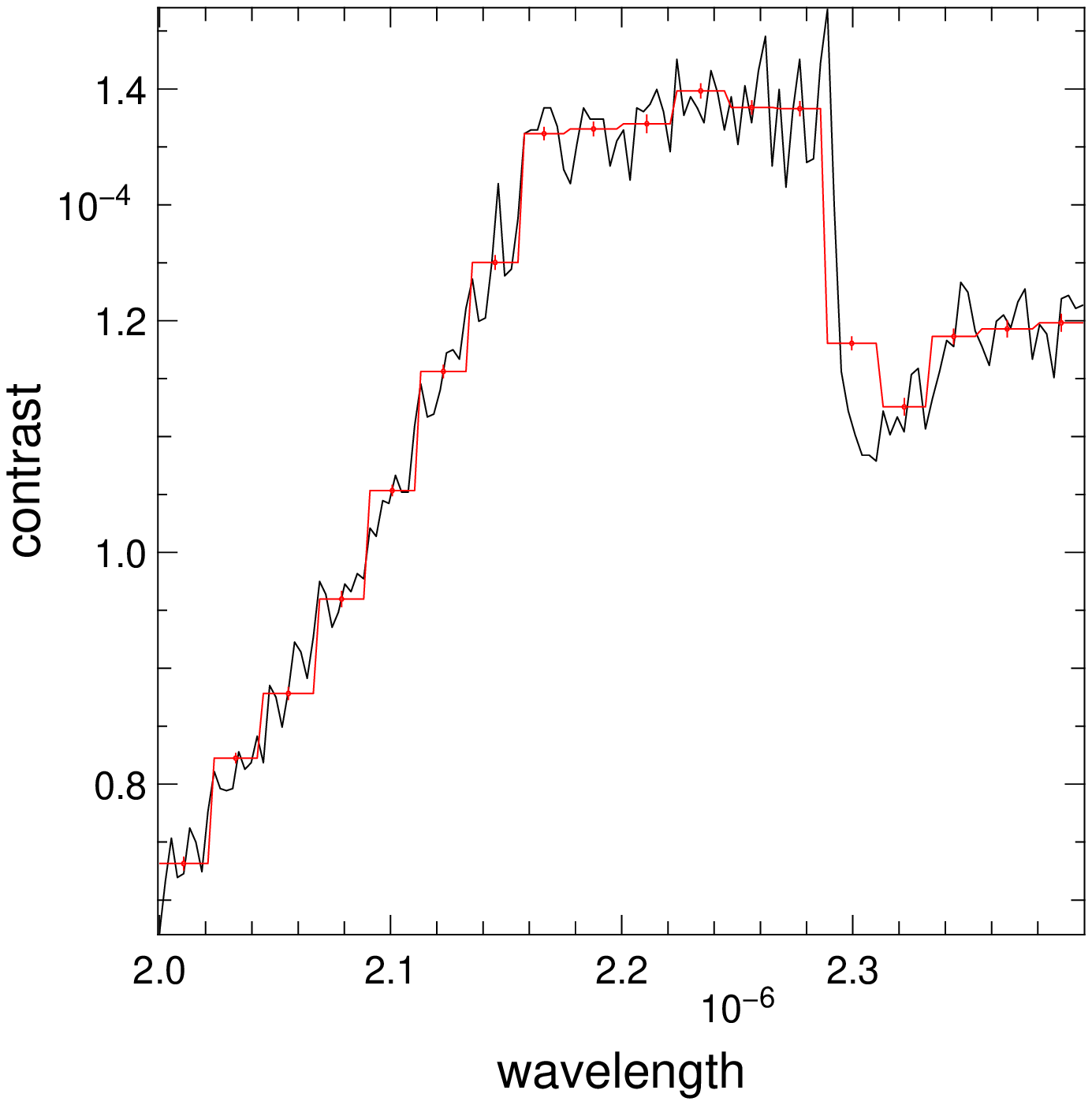}&
    \includegraphics[width=0.3\textwidth]{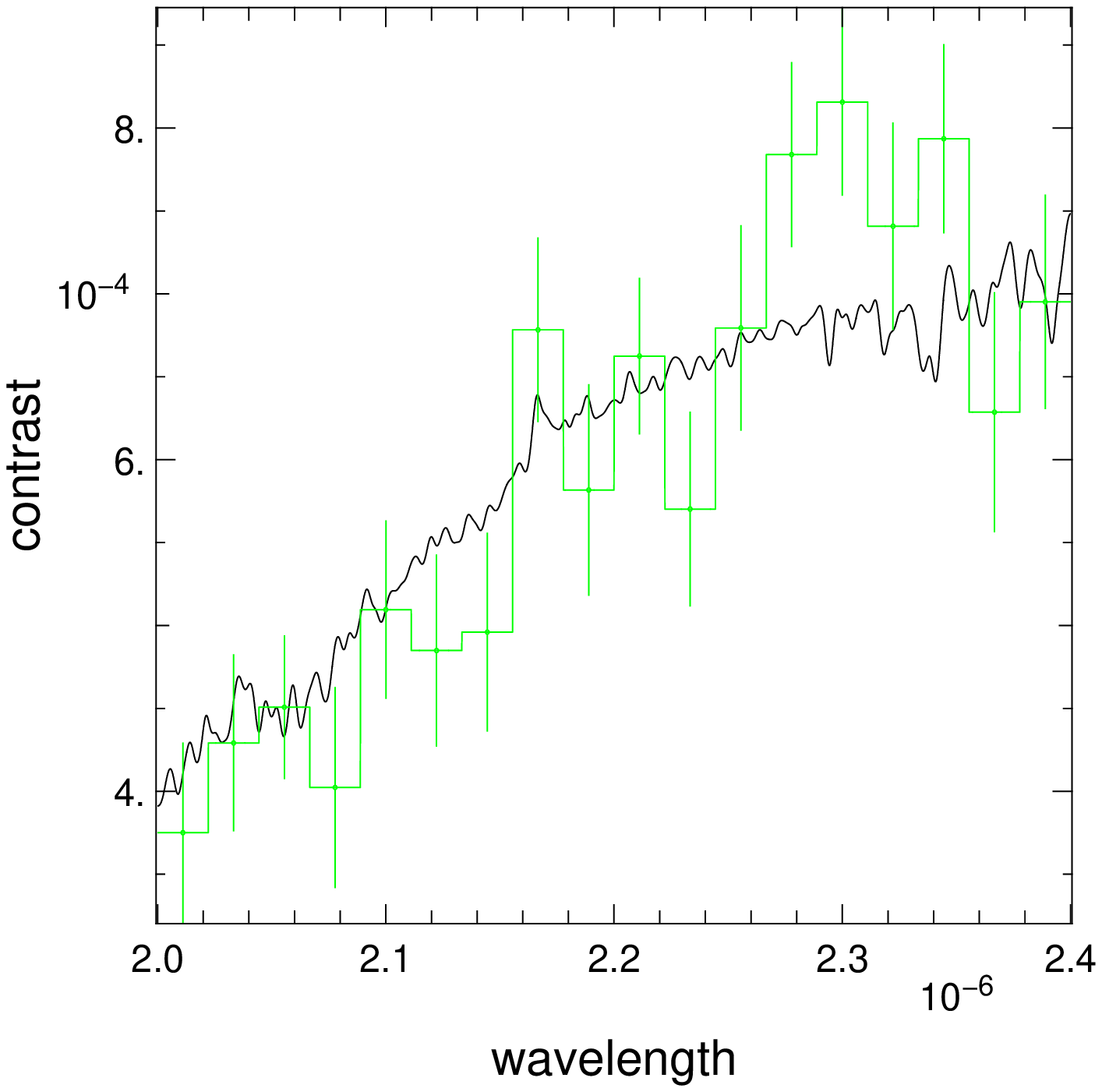}&
    \includegraphics[width=0.3\textwidth]{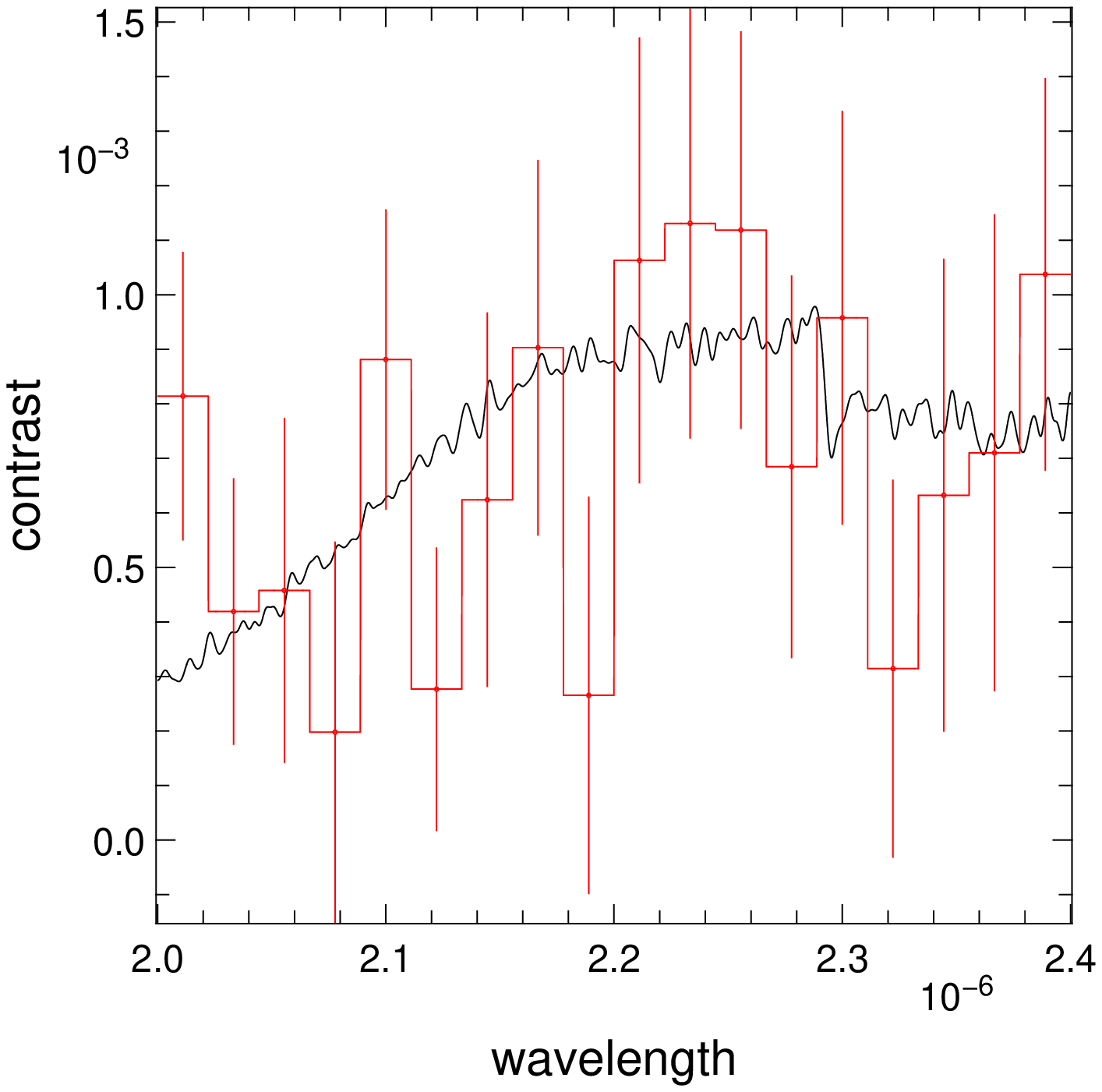}
    \\
    \includegraphics[width=0.3\textwidth]{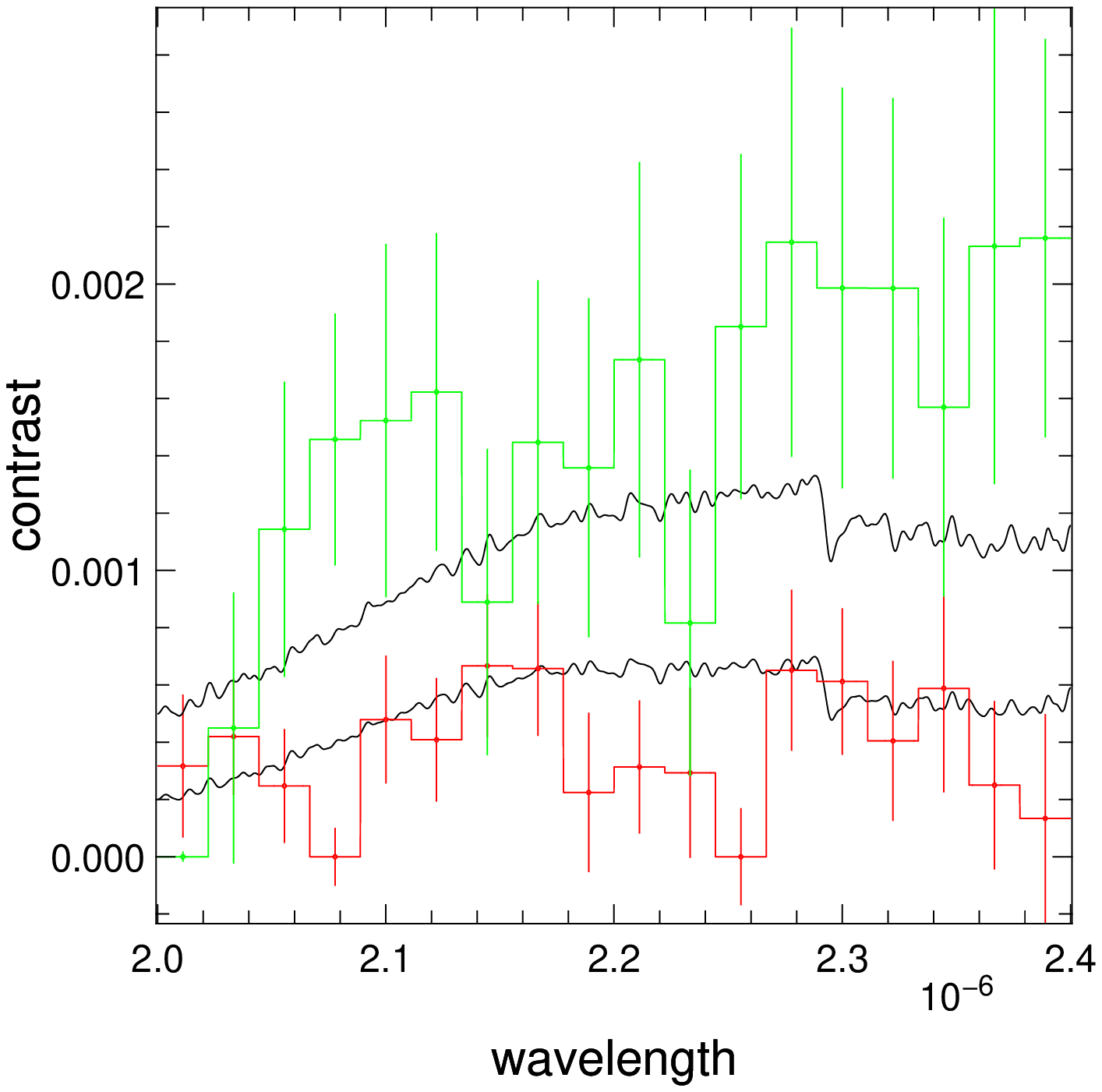}&
    \includegraphics[width=0.3\textwidth]{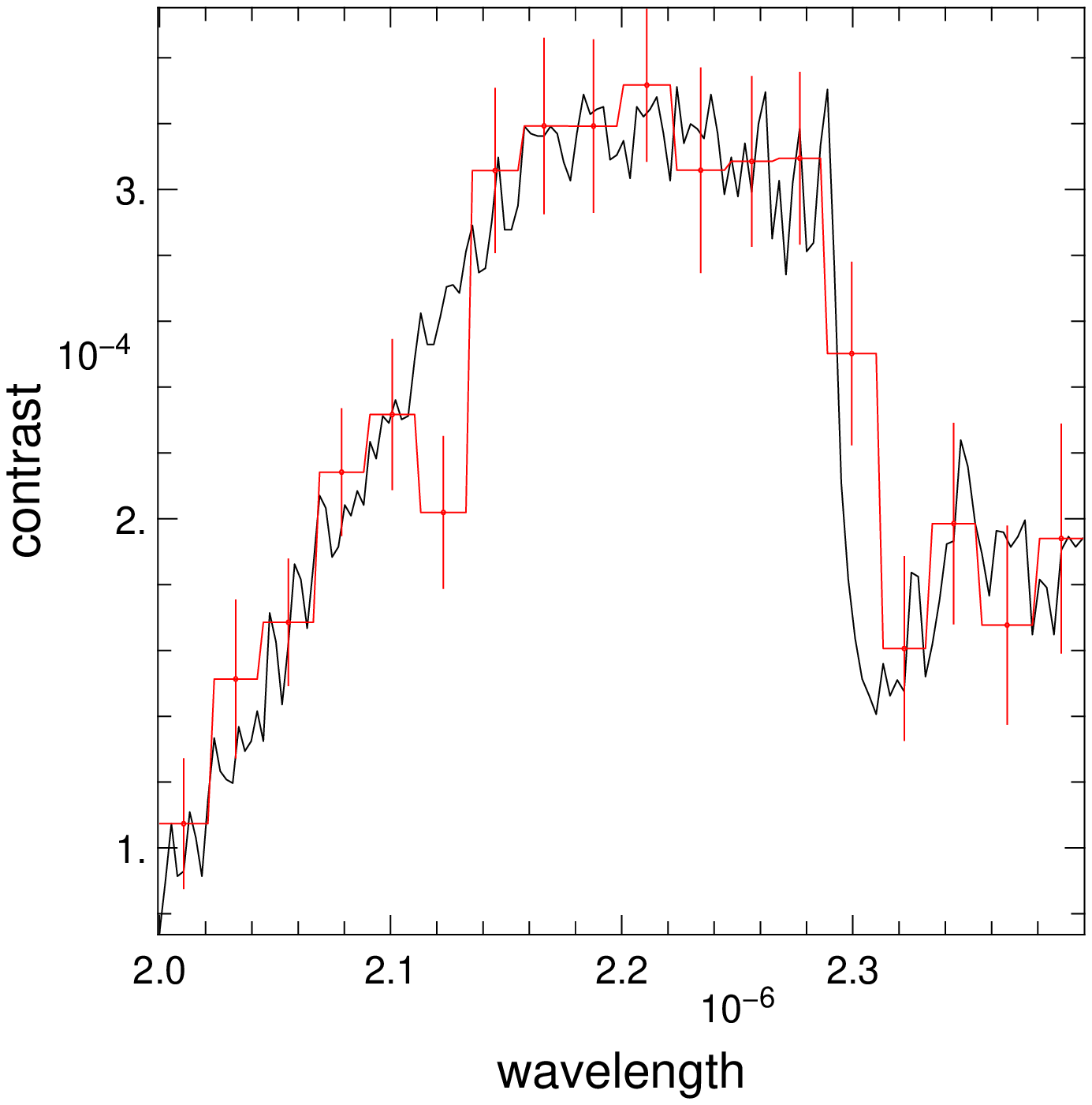}&
    \includegraphics[width=0.3\textwidth]{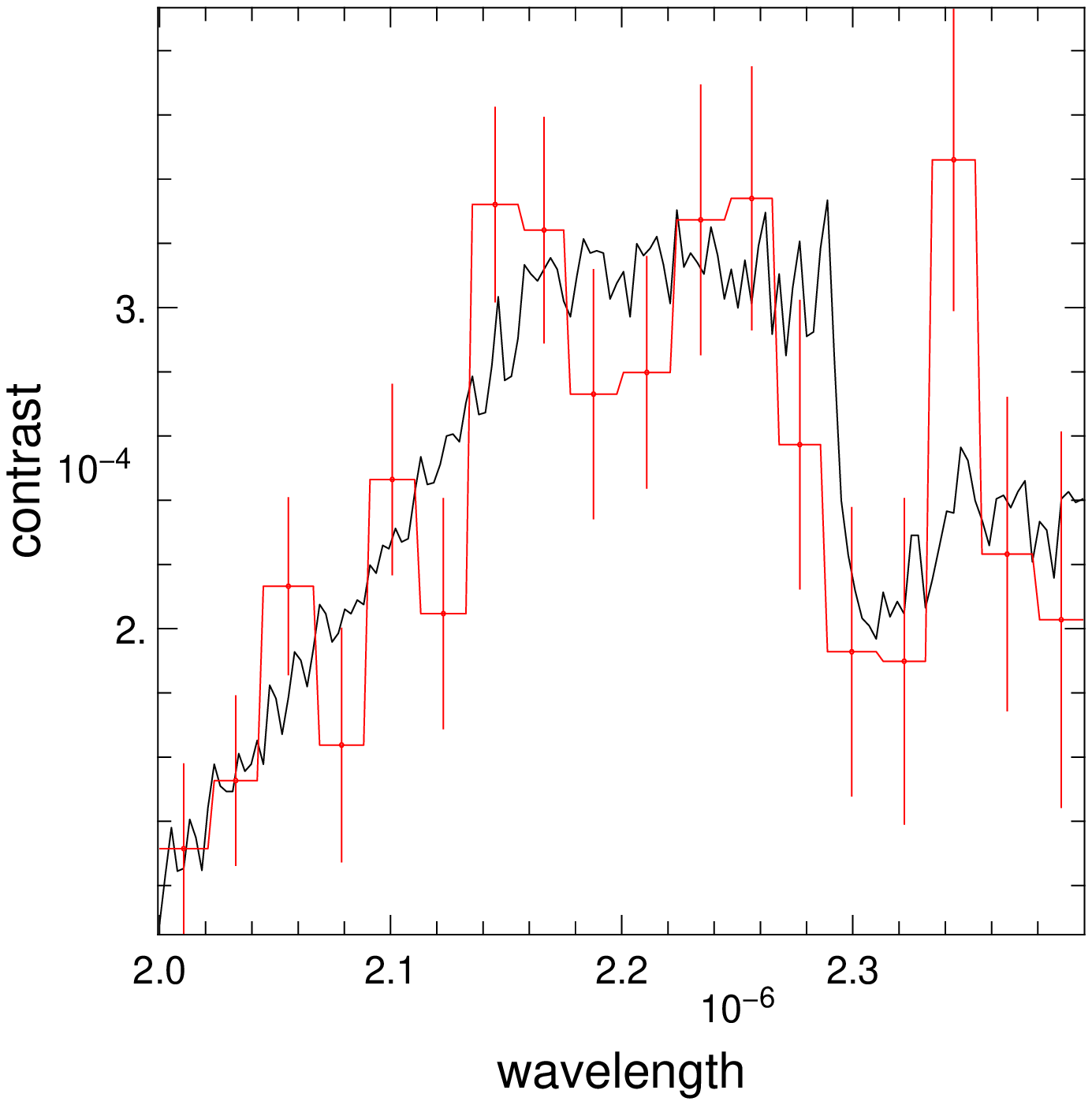}
  \end{tabular}
  \caption{Fit to the planet/star contrast from the simulated closure
    phase data in $K$ band (from left to right and top to bottom :
    $\tau$ Boo, HD 179949 b, HD 189733 b, HD 73256 b, 51 Peg b and HD
    209458 b). Red curves are used for the best-fit model when the
    input synthetic spectrum assumes heat redistribution around the
    whole planet, while green curves are used when the input spectrum
    assumes heat redistribution on the day side only.}
  \label{fig:spectre1}
\end{figure}

\begin{figure}[tp]
  \centering
  \begin{tabular}{ccc}
    \includegraphics[width=0.3\textwidth]{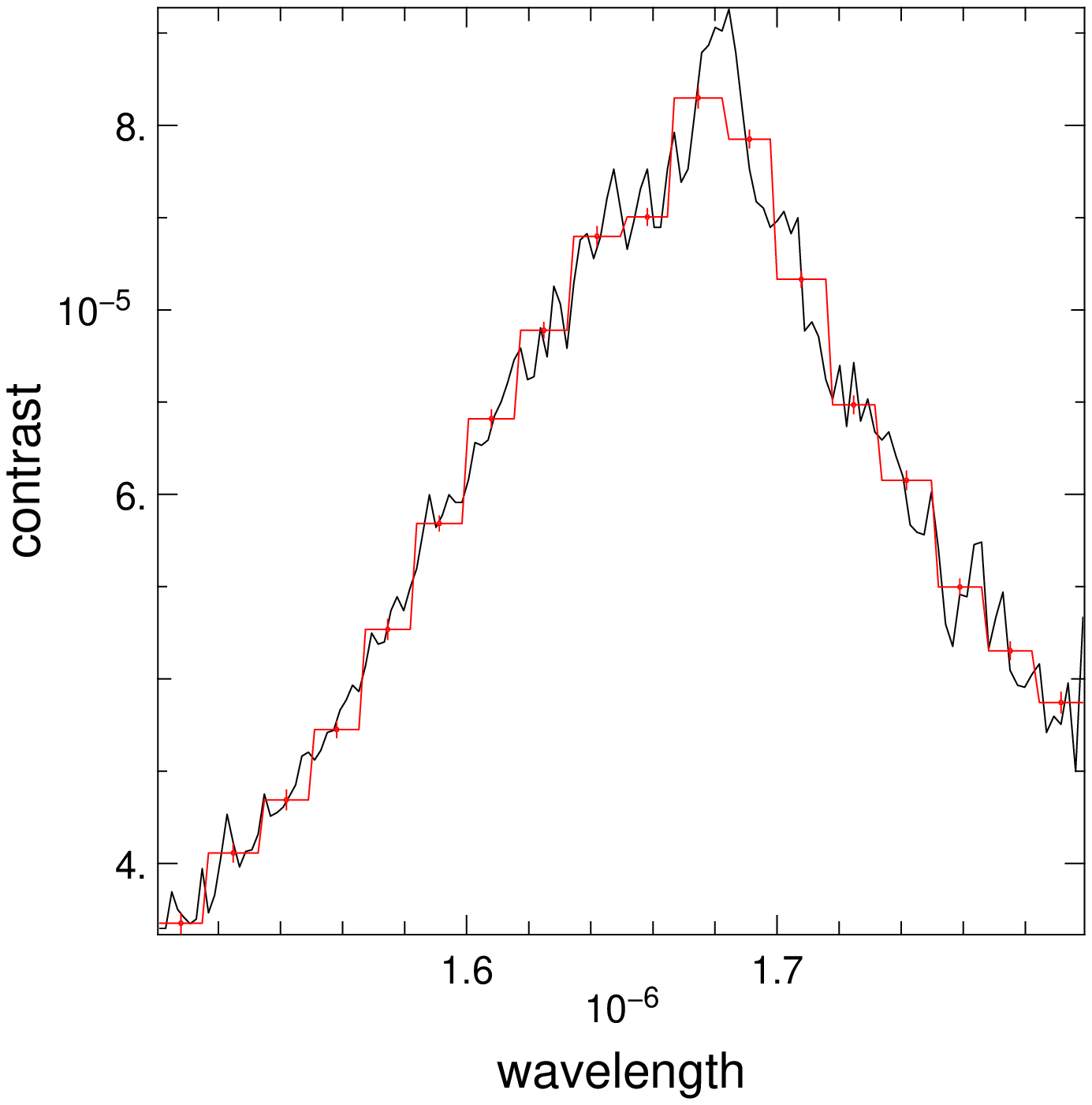}&
    \includegraphics[width=0.3\textwidth]{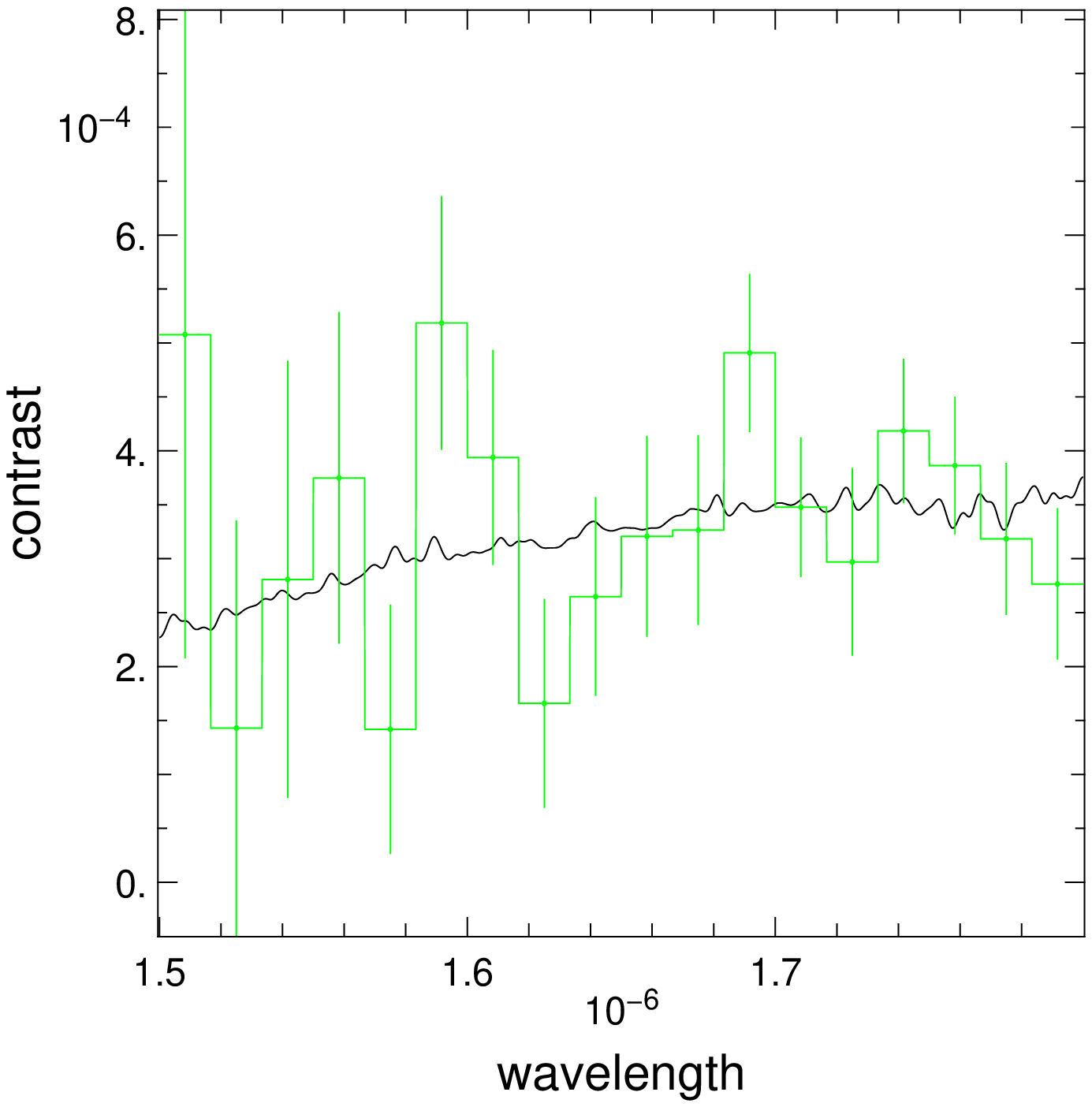}&
    \includegraphics[width=0.3\textwidth]{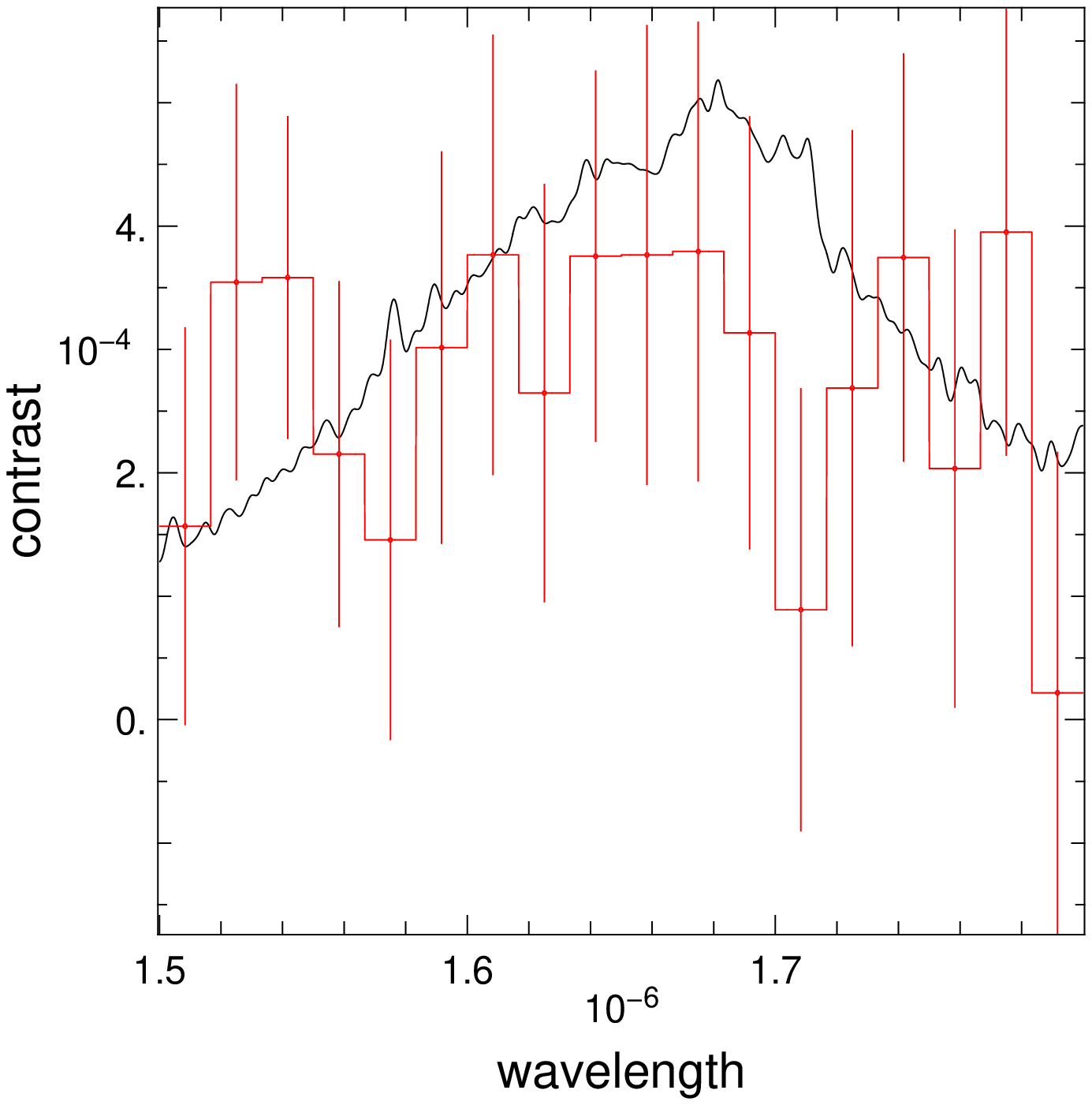}
    \\
    \includegraphics[width=0.3\textwidth]{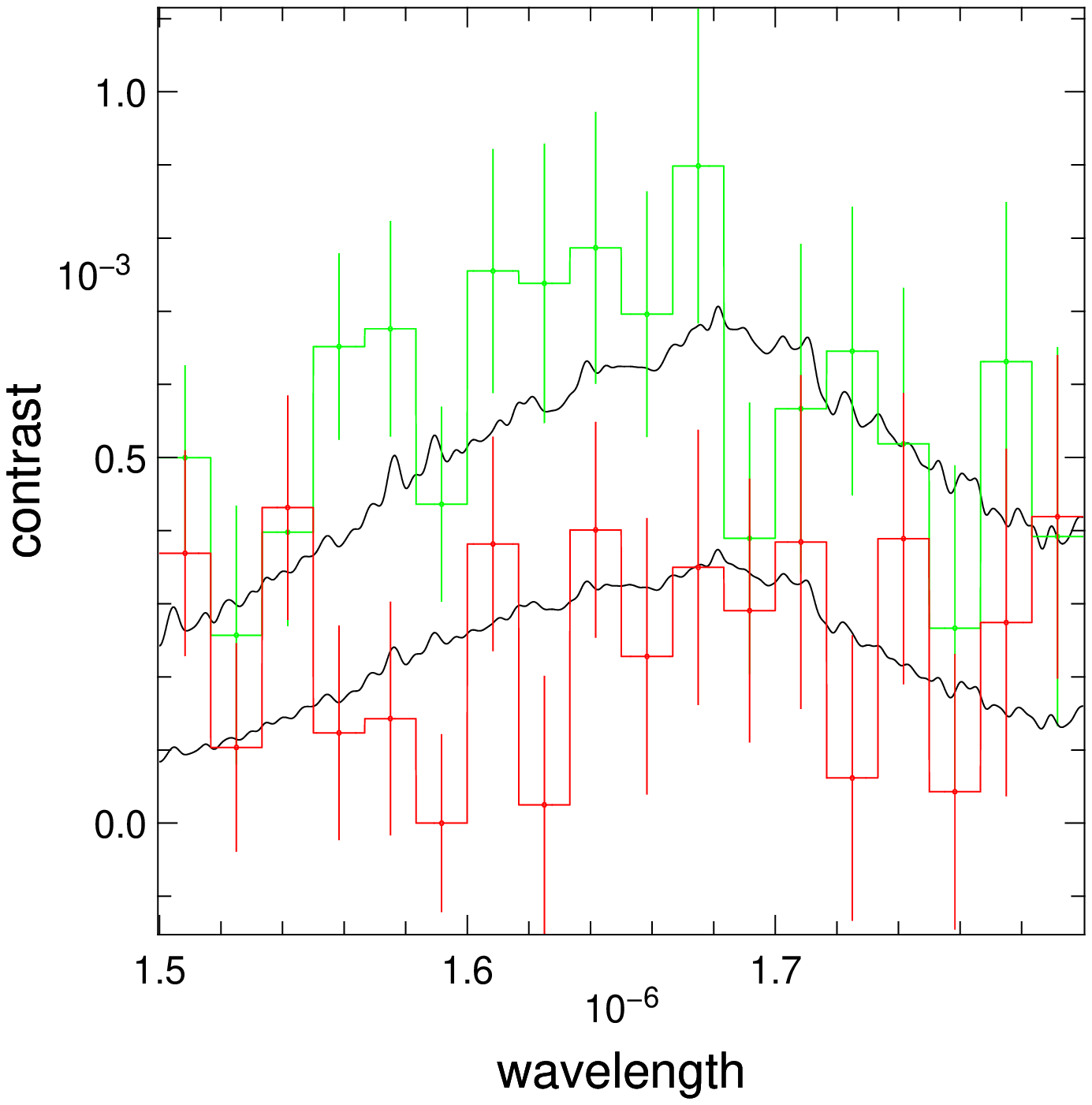}&
    \includegraphics[width=0.3\textwidth]{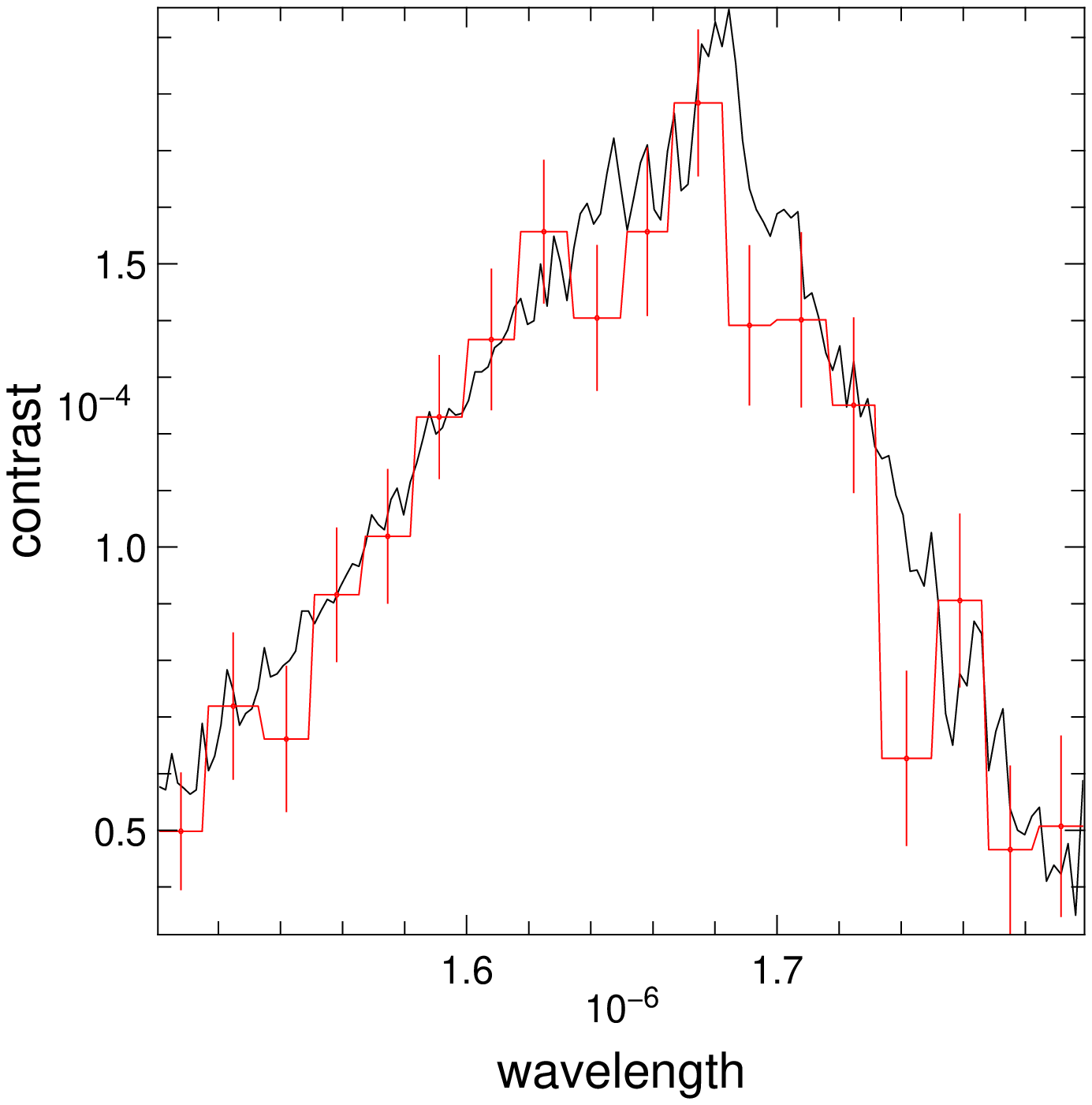}&
    \includegraphics[width=0.3\textwidth]{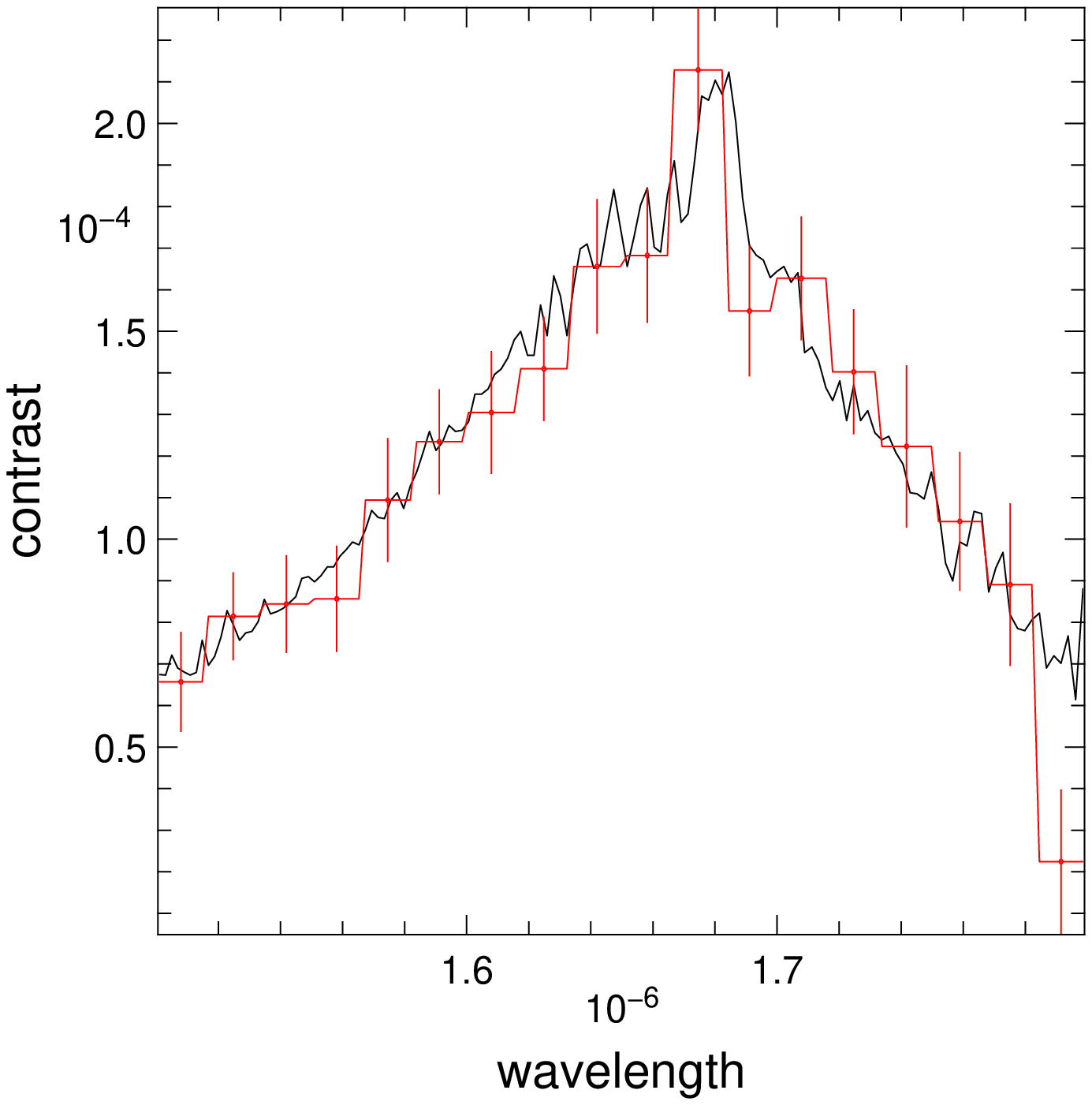}
  \end{tabular}
  \caption{Fit to the planet/star contrast from the simulated closure
    phase in the $H$ band (same conventions as in Fig.~\ref{fig:spectre1}).}
  \label{fig:spectre2}
\end{figure}

In a second step, we fix the best-fit orbital parameters obtained under
the black body assumption and perform a fit on the only planetary
radius individually for each spectral channel. In this step, we allow
the planetary radius to change across the various spectral channels,
using a black body assumption on each individual channel. The obtained
planetary radii are then converted to the value of the planet/star
contrast, which is the interesting quantity in this case. This fit is
illustrated in Fig.~\ref{fig:spectre1} for the $K$ band and in
Fig.~\ref{fig:spectre2} for the $H$ band.

\section{Discussion}
\label{sec:conclu}

From the results of the fit of the planetary spectra, it becomes
evident that VSI will be a powerful tool to characterize hot
EGPs. The simultaneous measurement of the contrast at various
wavelengths provides an insight into the thermal, physical and
dynamical structure of their atmospheres. In particular, the slope of
the spectrum in $H$ and $K$ band directly informs on the presence of
CH$_4$ in the planetary atmosphere, while the CO absorption feature
around 2.3 $\mu$m could also be detected from some of the selected
targets. Furthermore, the repeated observations at various orbital
phases provide an important information to constrain the heat
distribution mechanisms by measuring the temperature and atmospheric
composition around the planet. 

We note that the simulated observations are more successful and
constraining if the star-planet system is close to the observer ($ \leq 20
pc$). For such targets, the VLTI angular resolution well matches the
star-planet separation and the planet is bright enough to provide a
good signal-to-noise ratio. It is thus recommended to choose the closest systems as the
first targets of VSI. 

In the light of these results, one can safely conclude that the prime
criterion for the selection of additional targets is the magnitude of
the host star: it drives the signal to noise ratio on the closure
phases and therefore the quality of the fit of the planetary data. The
semi-major axis of the planetary orbit, or more precisely, the
temperature of the planetary companion, is of course another critical
parameter, as well as its radius (which is generally unknown). From these observations, additional targets can be proposed for
the VSI exoplanet sample such as HD 75286 b and HD 160691 d (from the
hot Jupiter family) or 55 Cnc e (a hot Neptune). All give
satisfactory results when repeating the above simulation
procedure, yet with larger relative error bars on the measured
planet/star contrast. A few other planets may be added to the list in
the coming years.

Systematic errors, not included in our simulations, could be a serious
limitation to these performance estimations. The use of integrated
optics is however expected to provide the required instrumental
stability (around $10^{-4}$) to enable the first thorough characterisation
of extrasolar planetary spectra in the near-infrared.

The direct detection of hot EGPs is undoubtely one of the most challenging VSI programs. This program was already one of AMBER's goal. However VSI intrinsic design offers multiple improvements with respect to AMBER which rely on two main axes : 
\begin{itemize}
 \item Improvement in the observable signal-to-noise ratio and accuracy : with a combining core made of an integrated optics circuit in which the incoming beams are spatially filtered out and routed so that each of the 4UT beams are carefully interfered with each other, the intrinsic stability of VSI is much higher than a classical bulk optics solution like AMBER. Moreover, unlike AMBER, VSI has included in its study an internal fringe tracker located as close as possible to the science instrument in order to control the stabilisation of the fringes and allow cophasing. This fringe tracker significantly increases the signal-to-noise ratio on the closure phases\cite{VSICorcione}.
 \item Increasing the number of simultaneous points: using 4 telescopes allows 4 closure phases to be measured simultaneously while AMBER  only permits one such measurement. This simultaneity reduces time dependant drifts, improves the calibration and permits to constrain the flux ratio by 4 points in a single measurements instead of 1 for AMBER, improving dramatically the quality of the fit.
\end{itemize}


\bibliography{SPIEposter_ok}   
\bibliographystyle{spiebib}   

\end{document}